# Two New Gradient Precondition Schemes for Full Waveform Inversion

Guanghui Huang    Huazhong Wang    Haoran Ren

( ① Institute of Computational Mathematics and Scientific/Engineering Computing;
② School of Ocean & Earth Science, Tongji University )

**Abstract** We propose two preconditioned gradient direction for full waveform inversion (FWI). The first one is using time integral wavefields. The Least square problem is formulated as the time integral residual wavefields, which can partially resolve the effect of high-passed filter in the traditional gradient formula; the convergence rate is greatly accelerated. The other one is localized offset Hessian inspired by the generalized imaging condition, which provides another redundancy in the Hessian. We compare the traditional conjugate gradient scaled by the shot illumination and localized offset Hessian (actually, only diagonal part is considered here), and contrast their performance for waveform inversion. The results demonstrate the localized offset Hessian (diagonal part) can provide much more information in the subsurface, and is preferred to the layer-strip inversion.

## 1 Introduction

Full waveform inversion is widely considered as a powerful tool for seismic imaging in the complex media, and is the most accurate parameter estimation method theoretically, and may be an alternative to the seismic migration in the future. However, due to the strong nonlinearity of FWI, severely sensitive to the initial guess model and source wavelet estimation, and a great heavy of computation, multiscale inversion strategy has been proposed in the geophysical exploration including time domain multigrid inversion (Bunks[1]) with successive filter technique from low to high frequency, frequency domain inversion with successive input individual frequency (Pratt[2], Sirgue, Pratt[3]), Laplace Fourier domain inversion with complex frequency parameter for damping high frequency component in the earlier inversion (Shin[4]). All of these strategies can be summarized as preconditioning gradient direction for updating wave-number from low to high, which can circumvent the cycle-skipping and improve the resolution of FWI. Another issue of FWI is that as we update the parameter with seismic waveform, due to the acquisition geometry limitation and geometrical spreading of seismic wave, deeper parts of subsurface model can't be well resolved and is updated quite slowly, as the contribution to the misfit function is small, even though the parameter error is big, hence, appropriate illumination compensation needs to be treated. Mathematically, this can be relaxed by solving so called Newton equation for correcting this effect. Due to the limitation of computation resource, it is unpractical to compute the element of full Newton Hessian matrix one by one. Ignoring the multiple scattering term, Gauss- Newton Hessian in the frequency domain is given by

$$H_a(x, y) = \sum_{\omega} \omega^4 \sum_{x_s} |f_s(\omega)|^2 \, G^*(x_s, x, \omega) G^*(x_s, y, \omega) \sum_{x_r} G(x, x_r, \omega) G(y, x_r, \omega)$$

It also costs a lot for computing the element of Gauss-Newton Hessian, as it needs forward modeling up to Nx*Nz, and can be reduced dramatically by considering the reciprocity of Green's function. Tang[5] developed the space domain Hessian calculation using phase-encoding techniques. This approach saves significant storage and computation time, but also introduces some crosstalk artifacts. Shin[6] proposes to estimate the diagonal of the Hessian via the so-called virtual sources, which is pointed out that they ignore the effect of receiver Green function, equivalently to be rewritten as (Mulder[7]),

$$\tilde{H}_a(x,x) = \sum_\omega \omega^4 \sum_{x_s} |f_s(\omega)|^2 |G^*(x_s,x,\omega)|^2 = \sum_\omega \omega^4 \sum_{x_s} |u_s(\omega)|^2 \quad (1)$$

This pseudo-diagonal Hessian can be implemented without any extra computation for amplitude preserving migration.

Plessix and Mulder[8] assume that the amplitude of receiver Green is simply proportional to the inverse of the distance between receiver and subsurface point by taking into account the limited coverage, and derive the following diagonal approximated Hessian,

$$\tilde{H}_a(x,x) = \sum_\omega \omega^4 \sum_{x_s} |u_s(\omega)|^2 \left[ \mathrm{asinh}(\frac{x_r^{max}(x_s)-x}{z}) - \mathrm{asinh}(\frac{x_r^{min}(x_s)-x}{z}) \right]$$

where $x^{min}$, $x^{max}$ is the minimum and maximum coverage within a single shot.

Hu et al. [9] recently proposes to use the inverse of approximate sparse Hessian matrix, which is constructed based on examining the auto-correlation and cross-correlation of the Jacobian matrix, which seems be a good alternative to the diagonal part of Hessian.

The paper is organized as follows. Firstly, we review the basic element of the full waveform inversion. And then, the first approach of circumventing the effect of high-passed filter in the gradient is proposed by formulating misfit functional based on integral wavefields. Next, localized offset Hessian matrix is given by generalizing Hessian based on generalized Born modeling, and we compare the diagonal part of Hessian matrix with Gauss-Newton Hessian. The contrast with performance of two preconditioner is to demonstrate the efficiency of our scheme, finally.

## 2 Integral Wavefields Misfit Functional

The generalized output least square (OLS) misfit functional defined as measuring the difference between observed and synthesized seismic waveform with data weighted operator is given by

$$J(v) = \frac{1}{2} \sum_{x_s} \sum_{x_r} \sum_t \| W_d (u(x_s,x_r,t) - d(x_s,x_r,t)) \|_2^2$$

Where $d$ is recorded waveform and $u$ is synthesized by constant-density acoustic equation.

For traditional FWI, the weighted data operator is identity operator $I$. And the gradient of misfit function with respect to the velocity is the zero-lag crosscorrelation between forward wavefields and reverse wavefields:

$$g = \frac{2}{v^3} \sum_{s=1}^{Ns} \int_0^T \frac{\partial^2 u_s}{\partial^2 t} v_s dt$$

where $u_s$ is the forward propagate wavefields and $v_s$ the back-propagated residual wavefields.

The gradient at the first iteration is equivalently to the approximated prestack Kirchhoff migration kinematically if the smooth initial velocity is used (Lailly, 1983) [10]. Such a gradient is high-passed filtered waveform caused by the second derivative of wavefields w.r.t time, appealing to migration for locating the structure, which contributes a lot of high wave number component to the velocity update. However, if the long wave-length velocity is not as accurate as enough, the high wavenumber update will give wrong velocity update, this is the cause of pour convergence of gradient-type algorithm for FWI and is our motivation for defining integral-type misfit function.

Our new objective function is defined as

$$J(v) = \frac{1}{2} \left\| \int_0^t (u_s(x,\tau) - d_s(x,\tau)) d\tau \right\|^2 \quad (2)$$

with data weighted operator $W_d u = \int_0^t u(x,\tau) d\tau$

We introduce the time integral wavefields $U(x,t)$ as

$$U(x,t) = \int_0^t u(x,\tau) d\tau$$

which satisfies the original acoustic wave equation, but propagates with integral source term:

$$\frac{1}{v^2} \frac{\partial^2 U}{\partial t^2} - \Delta U = \delta(\vec{x} - \vec{x}_s) \int_0^t f(\tau) d\tau$$

and $\int_0^t f(\tau) d\tau = H(t) * f(t)$ ($H(t)$ is Heaviside step function).

Therefore, $U(x,t)$ is the wavefields produced by low passed seismic wavelet. The gradient can be viewed as the low-passed filtered of traditional gradient. Actually, the gradient of newly proposed misfit can be derived by the adjoint-state technique (Lions [11]), and is given by

$$g = -\sum_{s=1}^{Ns}\int_0^T u_s v_s dt \qquad (3)$$

## 2.1 Example1

We test our full waveform inversion based on time integral misfit function in the 2D synthetic Marmousi model data. We generate the synthesized data at 80 shots with 1.2s data recording time length and full aperture receivers. The velocity of the starting model is linearly increasing with depth from 1.5km/s to 4.5km/s: v(z)=1500+z;

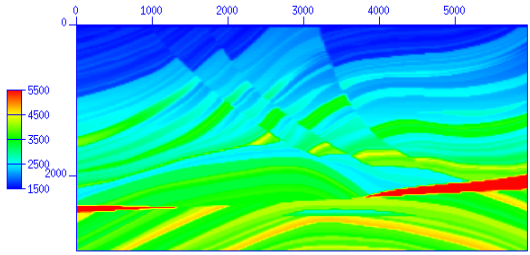

**Figure1 True Marmousi model**

The velocity model showed in Figure 2 (a), (b) is inverted after 200 iteration using time domain finite difference as forward modeling engine and limited BFGS as inversion strategy. The inversion results (b) is much more accurate than (a), especially for the deeper parts. Thanks to the well inverted low wave-number velocity at the early process. The convergence rate is also accelerated based on our time integral wavefields showed in the Figure 3. the 220-th and 240-th trace of velocity is extracted for comparing the accuracy of inverted velocity. The results confirm our aforementioned analysis. We believe that we can get much more accurate velocity inversion if we use layer-strip inversion strategy for improving deep parts of the subsurface velocity model.

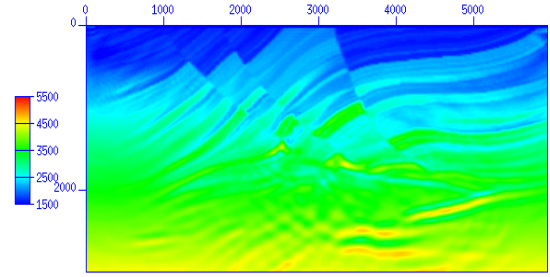

(a)

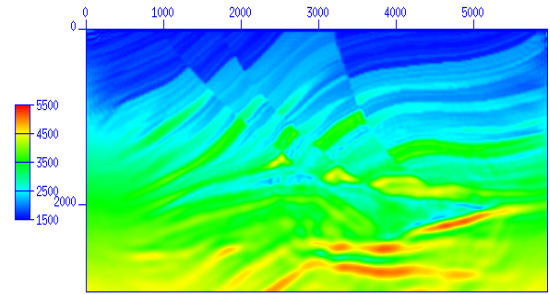

(b)

**Figure2 Inverted velocity after 200 iterations**

(a) conventional least square misfit function (b) time integral least square misfit function

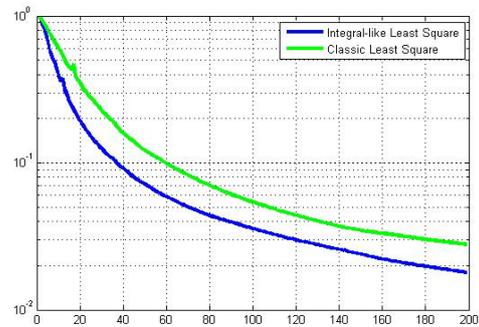

**Figure3 Misfit convergence history---logarithm map**

## 3 Sub-Surface Offset Hessian Preconditioner

For Gauss-Newton method in FWI, each iteration for updating velocity is equivalently to solve the following linearized least square problem:

$$\min_{\delta v} \frac{1}{2}\left\|(DF[v]\delta v - \delta d)\right\|_2^2$$

where $v$ is the background velocity, $\delta v$ is the velocity contrast, $\delta d$ is scattering wavefields received at the surface, and $DF[v]$ is Born modeling operator which is defined by

$$DF[v]\delta v = -\frac{2}{v^3}\int \omega^2 G(x,x_r,\omega)G(x_s,x,\omega)f_s(\omega)\delta v d^2 x \quad (4)$$

With the extension theory of model by Symes[12,13], we extend the born modeling operator as

$$\delta d(x_r,x_s,\omega) = -\frac{2}{v^3}\int \omega^2 G(x_r,x-h,\omega)G(x+h,x_s,\omega)f_s(\omega)\delta v d^2 x \quad (5)$$

The migration operator is the adjoint operator of extended Born modeling operator:

$$\overline{\delta v} = -\frac{2}{v^3}\int \omega^2 f_s^*(\omega)G^*(x+h,x_s,\omega)G^*(x_r,x-h,\omega)\delta d(x_r,x_s,\omega)dx_s dx_r d\omega \quad (6)$$

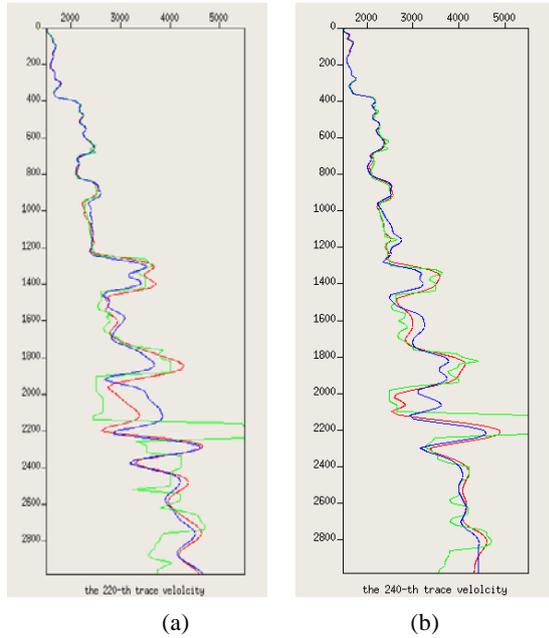

(a)        (b)

**Figure4 Inverted velocity comparison for single trace**

Green: the true model, blue: conventional functional red: new objective functional, (a) the 220-th trace velocity(b) the 240-th trace velocity

Hence, the sub-surface offset Hessian matrix can be access via substituting (5) into (6), then

$$H_a(x,h,y,\bar{h}) = \sum_\omega \frac{4\omega^4}{v^3(x)v^3(y)}\sum_{x_s}|f_s(\omega)|^2 G^*(x+h,x_s,\omega)G^*(y+\bar{h},x_s,\omega)\sum_{x_r}G(x_r,x-h,\omega)G(x_r,y-\bar{h},\omega)$$

The similar sub-surface Hessian matrix can is also obtained by Valenciano[14], who first considers generalized imaging condition in the ODCIGs, then derives the Hessian matrix in the sense of the least square.

However, our Hessian matrix includes the effect of the band-limited wavelet, which is quite important in the inversion and missing in his formulas.

However, the dimension of Hessian matrix is quite high, up to 8, if the horizon and vertical offset is introduced. The computation of the matrix is time-consuming and need huge disk storage. Actually, we use the diagonal part of matrix. As it can provide much information in the Hessian locally, due to the offset redundancy in the Hessian matrix.

### 3.1 Example2

We test the preconditioned conjugate gradient method with the same synthesized Marmousi model data as before mentioned above. For comparing the efficiency of improving the accuracy of velocity in the deep portion, we use the smoothed Marmousi model with Gaussian low-passed filter. The velocity models with preconditioned PRCG methods after 270 iterations are depicted in the Figure 5(b), (c) and (d). The deeper parts of the inverted velocity are significantly improved when using sub-offset diagonal Hessian matrix, which can be observed in the Figure5 (d). The convergence rate is also improved which is plotted in the Figure 6.

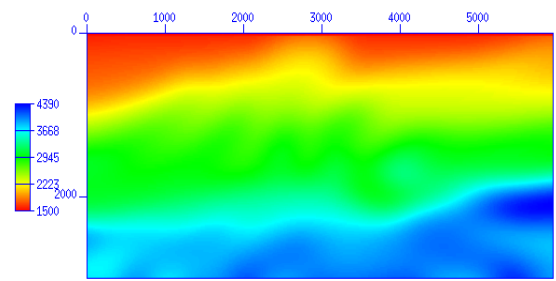

(a)

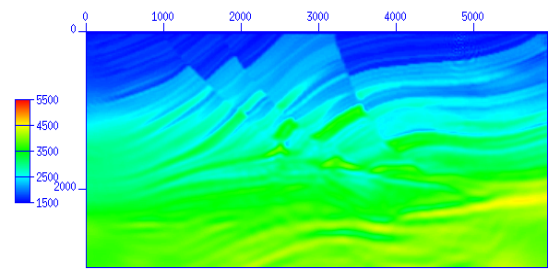

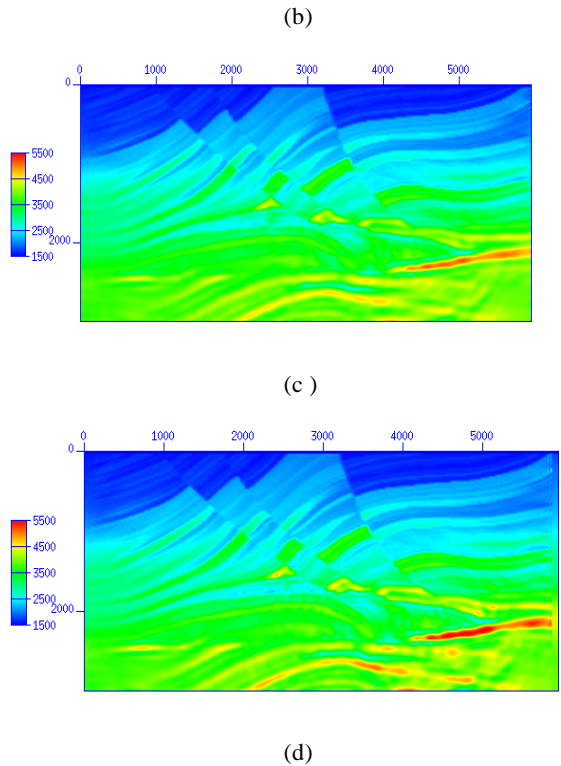

(b)

(c)

(d)

**Figure 5 Inverted velocity comparison preconditioned by different methods**

(a) Gaussian smoothed Marmousi initial model; other three are the inverted velocity model after 270 iterations using (b) unpreconditioned PRCG method (c) preconditioned PRCG by shot scaling (1) (d) preconditioned by diagonal sub-offset Hessian PRCG

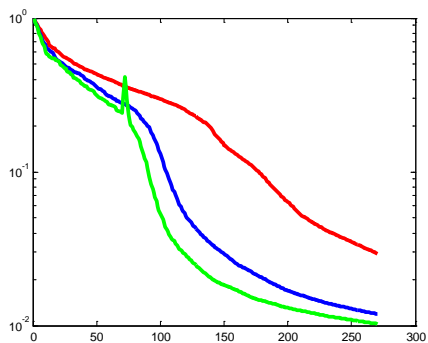

**Figure 6 Red: unpreconditioned PRCG; Blue: shot scaling Green: sub-offset Hessian scaling.**

## 4 Conclusions

The true reflectivity model is actually blurred by the Hessian operator. This operator can be roughly decompositioned into two multiplicity operator: data band-limited and limited aperture operator. The time integral residual plays the role in the low-passed filter of residual wavefields, and make the gradient with much more lower wavenumber, thus, this can improve resolution of inversion. Sub-offset Hessian matrix is a good scaling of gradient for improving the resolution of deeper portion of velocity model. However, how to extract the much more convolution information from this big Hessian matrix, while not bringing a great heavy of computation will be investigated in the future research.

**Acknowledgements**

The authors grateful acknowledge the financial support by "973" project (2011CB202402) and also thank the financial support by the great and special project (2008ZX05005-005-007HZ) and (2008ZX05023-005-016). Special thanks give to Sinopec Geophysical Research Institute (SGRI) and Geophysics Research Institute of Shengli Oilfield.